\documentstyle[12pt,epsfig]{article}
\def\lbldef#1#2{\expandafter\gdef\csname #1\endcsname {#2}}

\def\href#1#2{#2}  
\begin{document}
\baselineskip=15.5pt
\pagestyle{plain}
\setcounter{page}{1}
%\renewcommand{\thefootnote}{\fnsymbol{footnote}}
%--------+---------+---------+---------+---------+---------+---------+
%Title page

\begin{titlepage}

\begin{flushright}
CERN-TH/2000-086\\
hep-th/0003174
\end{flushright}
\vspace{10 mm}

\begin{center}
{\Large Dilatonic $p$-Branes and Brane Worlds}

\vspace{5mm}

\end{center}

\vspace{5 mm}

\begin{center}
{\large Donam Youm\footnote{E-mail: Donam.Youm@cern.ch}}

\vspace{3mm}

Theory Division, CERN, CH-1211, Geneva 23, Switzerland

\end{center}

\vspace{1cm}

\begin{center}
{\large Abstract}
\end{center}

\noindent

We study a general dilatonic $p$-brane solution in arbitrary dimensions in 
relation to the Randall-Sundrum scenario.  When the $p$-brane is fully 
localized along its transverse directions, the Kaluza-Klein zero mode of 
bulk graviton is not normalizable.  When the $p$-brane is delocalized along 
its transverse directions except one, the Kaluza-Klein zero mode of bulk 
graviton is normalizable if the warp factor is chosen to increase, in which 
case there are singularities at finite distance away from the $p$-brane.  
Such delocalized $p$-brane can be regarded as a dilatonic domain wall as seen 
in higher dimensions.  This unusual property of the warp factor allows one to 
avoid a problem of dilatonic domain wall with decreasing warp factor that 
free massive particles are repelled from the domain wall and hit 
singularities, since massive particles with finite energy are trapped around 
delocalized $p$-branes with increasing warp factor by gravitational force and 
can never reach the singularities.  

\vspace{1cm}
\begin{flushleft}
CERN-TH/2000-086\\
March, 2000
\end{flushleft}
\end{titlepage}
\newpage

\section{Introduction}

According to the brane world scenario \cite{add,aadd}, the extra spatial 
dimensions can be as large as a millimeter without contradicting the current 
experimental observations, if the fields of Standard Model are confined 
within a brane.  More recently proposed scenario by Randall and Sundrum 
(RS) \cite{ran1,ran2,ran3} even allows infinite extra space because of 
the special property of warped spacetime that leads to localization of 
gravity around domain wall.  So, the brane world scenarios open up the 
possibility of probing the extra dimensions in the near future.  Furthermore, 
the brane world scenarios provide with a new framework for solving the 
hierarchy problem of particle physics.

In order for the RS brane world scenario to describe our world, it 
has to be derivable from well-established theories.  The previous 
studies attempt to embed the RS domain wall (with the exponentially 
decreasing warp factor) into supergravity theories 
\cite{bc,st,cg,kls,bbs,clp,kl,bc2,bbs2,abn} or domain walls which localize 
gravity into string theories \cite{bs,youm1}.  
Also, a main motivation for the embedding into supergravity theories 
is that the fine-tuned value of domain wall tension in terms of the bulk 
cosmological constant is required by supersymmetry through the BPS 
condition, if such embedding is possible.  

In this paper, we study the Kaluza-Klein (KK) zero mode of graviton in 
the bulk of a general dilatonic $p$-brane in arbitrary dimensions for the 
purpose of seeing any relevance to the RS type scenario.  Since the 
brane world scenarios assume that fields of Standard Model are identified 
as the worldvolume fields of $p$-branes in string theories, it is natural 
to embed domain walls of the RS type models into $p$-branes in higher 
dimensions.  So, it appears that the embedding into $p$-branes in 
string theories is a natural way of realizing the RS type scenario 
within string theories.

We find in general that when a $p$-brane is fully localized along its 
transverse directions, the KK zero mode of bulk graviton cannot be 
normalized, thereby the $(p+1)$-dimensional gravity in the worldvolume 
of the $p$-brane cannot be realized through the KK zero mode.  
However, if the $p$-brane is delocalized along its transverse directions 
except one, the KK zero mode of bulk graviton is normalizable.  This implies 
that if our four-dimensional world is embedded in the worldvolume of some 
(intersecting) $p$-brane of string theories through the RS type scenario, 
then some of the transverse directions of the brane has to be compact or the 
range of the radial coordinate of the transverse space has to be within a 
finite interval.  Unlike the case of domain walls (with codimension one), 
the KK zero mode of bulk graviton is normalizable if the warp factor 
increases, in which case there are singularities at finite distance from the 
brane.  Due to the increasing warp factor, free massive particles are 
attracted towards the brane, never hitting the singularities, in contrast to 
the case of domain walls with decreasing warp factor, which repels free 
massive particles \cite{mvv,grs}.  So, a delocalized $p$-brane with 
increasing warp factor does not suffer from the problem of a domain wall 
with decreasing warp factor that massive matter escapes into the extra 
spatial dimensions due to the gravitational repulsion by the wall.

The paper is organized as follows.  In section 2, we summarize the dilatonic 
domain wall solution in relation to the RS scenario.  We study the KK zero 
mode of graviton in the bulk of fully localized dilatonic $p$-brane in 
section 3 and in the bulk of delocalized dilatonic $p$-brane in section 4.

\section{Dilatonic Domain Wall Solution}

In this section, we summarize a general dilatonic domain wall solution  
in arbitrary dimensions for the purpose of reference in the later 
sections, where we will make comparison of the properties of such domain 
wall solutions to those of $p$-brane solutions.  The Einstein frame action 
for the domain wall solution has the following form:
\begin{equation}
S_{\rm DW}={1\over{2\kappa^2_D}}\int d^Dx\sqrt{-g}\left[{\cal R}_g
-{4\over{D-2}}\partial_{\mu}\phi\partial^{\mu}\phi+e^{-2a\phi}\Lambda\right].
\label{dwact}
\end{equation}
The extreme dilatonic domain wall solution is given by:
\begin{eqnarray}
g_{\mu\nu}dx^{\mu}dx^{\nu}&=&H^{4\over{(D-2)\Delta}}\left[-dt^2+dx^2_1+\cdots+
dx^2_{D-2}\right]+H^{{4(D-1)}\over{(D-2)\Delta}}dx^2_{D-1},
\cr
e^{2\phi}&=&H^{{(D-2)a}\over{\Delta}},\ \ \ \ \ \ 
\Delta\equiv{{(D-2)a^2}\over{2}}-{{2(D-1)}\over{D-2}},
\label{dwsol}
\end{eqnarray}
where the harmonic function $H$ for the domain wall located at $x_{D-1}=0$ is
\begin{equation}
H=1+Q|x_{D-1}|=1\pm\sqrt{-{{\Delta\Lambda}\over{2}}}|x_{D-1}|,
\label{dwharmf}
\end{equation}
where the invariance under the ${\bf Z}_2$ transformation $x_{D-1}\to 
-x_{D-1}$ is imposed.  By redefining the transverse coordinate $x_{D-1}$, one 
can put the domain wall metric into the following standard form of the RS 
brane world scenario:
\begin{equation}
g_{\mu\nu}dx^{\mu}dx^{\nu}=e^{2A(y)}\left[-dt^2+dx^2_1+\cdots+dx^2_{D-2}
\right]+dy^2,
\label{warpdwmet}
\end{equation}
where the warp factor $e^{2A}$ and the dilaton $\phi$ are given by
\begin{equation}
e^{2A}=(K|y|+1)^{8\over{(D-2)^2a^2}},\ \ \ \ \ \ 
\phi={1\over a}\ln(K|y|+1),
\label{warpdil}
\end{equation}
where 
\begin{equation}
K=\pm{{(D-2)a^2}\over 2}\sqrt{-{\Lambda\over{2\Delta}}}.
\label{defk}
\end{equation}
It is observed \cite{youm1,stfdw,gjs,youm3} that when the sign $\pm$ in the 
above expression for $K$ is chosen so that the warp factor $e^{2A}$ has 
a root at a finite non-zero value of $y$ on both sides of the domain wall, 
there exists the normalizable KK zero mode graviton bound state, which can 
be identified as a massless graviton in one lower dimensions.  In such 
case, the warp factor decreases on both sides of the wall, so the RS 
type scenario can be realized.  However, a problem with such domain wall is 
that the roots of the warp factor correspond to naked singularities, which 
are undesirable unless significant physical meanings are associated with 
them.   The resolution or the physical interpretation of such naked 
singularities is still an open question.  With different choice of sign 
$\pm$ in Eq. (\ref{defk}), there is no singularity at finite nonzero $y$, 
but the normalizable graviton KK zero mode does not exist.  

\section{Fully Localized $p$-Branes}

We begin by summarizing a general dilatonic $p$-brane solution in $D$ 
spacetime dimensions, where $p\geq 3$ and $D>5$.  The reason for considering 
such solution is that it covers all the possible single charged $p$-branes 
in string theories.  In this paper, we regard the $p$-brane as a solitonic 
brane magnetically charged under the field strength $F_n$ of the rank 
$n\equiv D-p-2$, for the reason which will become clear in the following.  
The action has the following form:
\begin{equation}
S_p={1\over{2\kappa^2_D}}\int d^Dx\sqrt{-G}\left[{\cal R}_G-{4\over{D-2}}
(\partial\phi)^2-{1\over{2\cdot n!}}e^{-2a_p\phi}F^2_n\right],
\label{pbrnact}
\end{equation}
where once again $D=p+n+2$.  The solution to the equations of motion of this 
action for the extreme dilatonic $p$-brane with the longitudinal coordinates 
${\bf x}=(x_1,...,x_p)$ and the transverse coordinates ${\bf y}=
(y_1,...,y_{n+1})$, located at ${\bf y}={\bf 0}$, has the following form:
\begin{eqnarray}
G_{MN}dx^Mdx^N&=&H^{-{{4(n-1)}\over{(p+n)\Delta_p}}}_p\left[-dt^2+d{\bf x}
\cdot d{\bf x}\right]+H^{{4(p+1)}\over{(p+n)\Delta_p}}_pd{\bf y}\cdot d{\bf y},
\cr
e^{2\phi}&=&H^{{(p+n)a_p}\over{\Delta_p}}_p,\ \ \ \ \ 
F_n=\star(dH_p\wedge dt\wedge dx_1\wedge\cdots\wedge dx_p),
\label{magpbrnsol}
\end{eqnarray}
where
\begin{equation}
H_p=1+{{Q_p}\over{|{\bf y}|^{n-1}}},\ \ \ \ \ \ 
\Delta_p={{(p+n)a^2_p}\over{2}}+{{2(p+1)(n-1)}\over{p+n}}.
\label{magpbrndef}
\end{equation}
We will consider $p$-branes with asymptotically flat spacetime, only, i.e. 
the $n>1$ case.  So, $\Delta_p$ defined in the above is always positive.  

To study the KK modes of graviton in the $p$-brane background, we consider 
the leading order Einstein's equations satisfied by the small fluctuations 
$h_{\mu\nu}(x^{\mu},{\bf y})$ around the $p$-brane metric.  In general, 
the linearized Einstein's equations $\delta{\cal G}_{MN}=\kappa^2_D\delta 
T_{MN}$ for the following form of the metric fluctuations
\begin{equation}
G_{MN}dx^Mdx^N=e^{B(y^k)}\left[\eta_{\mu\nu}+h_{\mu\nu}(x^{\rho},y^k)
\right]dx^{\mu}dx^{\nu}+g_{ij}(y^k)dy^idy^j,
\label{metrcfluc}
\end{equation}
is given, in the transverse traceless gauge $h^{\mu}_{\ \mu}=0=\partial^{\mu}
h_{\mu\nu}$, by (Cf. Ref. \cite{cceh})
\begin{equation}
e^{-B}\eta^{\rho\sigma}\partial_{\rho}\partial_{\sigma}h_{\mu\nu}+
e^{-{{p+1}\over 2}B}{1\over\sqrt{g}}\partial_i\left[e^{{{p+1}\over 2}B}
\sqrt{g}g^{ij}\partial_jh_{\mu\nu}\right]=0,
\label{lineineqn}
\end{equation}
if the stress tensor $T_{MN}$ satisfies the following condition:
\begin{equation}
\delta T_{MN}=T^P_Mh_{PN}.
\label{tensorcond}
\end{equation}
Here, $\delta{\cal G}_{MN}$ and $\delta T_{MN}$ denote the variation 
of the Einstein tensor ${\cal G}_{MN}$ and the stress tensor $T_{MN}$ with 
respect to the metric perturbation $\eta_{\mu\nu}\to\eta_{\mu\nu}+
h_{\mu\nu}$.  Of course, in Eq. (\ref{tensorcond}) it is assumed that 
$h_{y^iy^j}=0$.   

It can be easily shown that the stress tensor condition (\ref{tensorcond}) 
is satisfied for the $p$-brane solution in Eq. (\ref{magpbrnsol}) by 
using the fact that the dilaton field $\phi$ depends only on the 
transverse coordinates ${\bf y}$ and only the transverse components 
of the form field strength $F_n$ are nonzero.  So, the linearized 
Einstein's equations in the transverse traceless gauge satisfied by the 
metric perturbation $h_{\mu\nu}$ around the $p$-brane metric 
(\ref{magpbrnsol}) of the following form:
\begin{equation}
G_{MN}dx^Mdx^N=H^{-{{4(n-1)}\over{(p+n)\Delta_p}}}_p\left[\eta_{\mu\nu}+
h_{\mu\nu}\right]dx^{\mu}dx^{\nu}+H^{{4(p+1)}\over{(p+n)\Delta_p}}_pd{\bf y}
\cdot d{\bf y}
\label{pbrnpert}
\end{equation}
is given by
\begin{equation}
\eta^{\rho\sigma}\partial_{\rho}\partial_{\sigma}h_{\mu\nu}+
H^{-{4\over{\Delta_p}}}_p\delta^{ij}\partial_i\partial_j
h_{\mu\nu}=0.
\label{pbranlinein}
\end{equation}
To consider the graviton KK mode of mass $m_{\alpha}$, we decompose the 
metric perturbation $h_{\mu\nu}$ as $h_{\mu\nu}(x^{\rho},y^k)=
\hat{h}^{(\alpha)}_{\mu\nu}(x^{\rho})f_{\alpha}(y^k)$ and require 
$\hat{h}^{(\alpha)}_{\mu\nu}$ to satisfy $\Box_x\hat{h}^{(\alpha)}_{\mu\nu}=
m^2_{\alpha}\hat{h}^{(\alpha)}_{\mu\nu}$, where $\Box_x\equiv\eta^{\rho\sigma}
\partial_{\rho}\partial_{\sigma}$.  Then, the linearized Einstein's 
equations (\ref{pbranlinein}) take the following form:
\begin{equation}
\nabla^2_yf_{\alpha}+m^2_{\alpha}H^{4\over{\Delta_p}}_pf_{\alpha}=0,
\label{kklineineq}
\end{equation}
where $\nabla^2_y\equiv \delta^{ij}\partial_i\partial_j$. 
We decompose the KK mode $f_{\alpha}({\bf y})$ into the radial and the 
angular parts as $f_{\alpha}({\bf y})=g_{\alpha}(y)Y_{\ell}(\Omega_n)$, 
where $y\equiv |{\bf y}|$ and $\Omega_n$ collectively denotes the angular 
coordinates of the unit $n$-sphere $S^n$.  The $n$-dimensional spherical 
harmonics $Y_{\ell}$ satisfies $\nabla^2_yY_{\ell}={{\ell(\ell+n-1)}\over{y^2}}
Y_{\ell}$.  So, Eq. (\ref{kklineineq}) reduces to the following form:
\begin{equation}
\partial_y\left[y^n\partial_yg_{\alpha}\right]+\ell(\ell+n-1)y^{n-2}
g_{\alpha}+m^2_{\alpha}y^nH^{4\over{\Delta_p}}_pg_{\alpha}=0,
\label{radlineineq}
\end{equation}
from which we see that the KK modes $g_{\alpha}$ with different masses 
$m_{\alpha}$ are orthogonalized with respect to the weighting function 
$w(y)=y^nH^{4\over{\Delta_p}}_p$.  From Eq. (\ref{kklineineq}), one 
can see that $f_0({\bf y})={\rm constant}$ is the KK zero mode ($m_0=0$).  
The normalization integral for the KK zero mode $g_0(y)={\rm constant}$ 
is as follows:
\begin{eqnarray}
\int^{\infty}_0dy\,y^nH^{4\over{\Delta_p}}_pg^2_0&=&
{{g^2_0Q^{4\over\Delta_p}_p}\over{(n-1){4\over\Delta_p}-(n+1)}}
y^{n+1-(n-1){4\over\Delta_p}}
\cr
& &\left.\times\,_2F_1[{{n+1}\over{n-1}}-{4\over\Delta_p},
-{4\over\Delta_p};{{2n}\over{n-1}}-{4\over\Delta_p};-{y^{n-1}\over Q_p}]
\right|^{y=\infty}_{y=0},
\label{kkznormint}
\end{eqnarray}
where $_2F_1$ is the hypergeometric function defined in terms of series as
\begin{equation}
_2F_1[a,b;c;z]=\sum^{\infty}_{l=0}{{(a)_l(b)_l}\over{(c)_l}}{{z^l}\over{l!}}=
1+{{ab}\over{c}}{z\over{1!}}+{{a(a+1)b(b+1)}\over{c(c+1)}}{{z^2}\over{2!}}
+\cdots.
\label{hypergeo}
\end{equation}
Note, this graviton KK zero mode normalization integral is equivalent to 
the normalization condition $\int d^{n+1}{\bf y}\,G^{tt}\sqrt{-G}
=V_n\int dy\,y^nH^{4\over{\Delta_p}}_p<\infty$, where $V_n$ is the volume 
of $S^n$, obtained in Ref. \cite{cceh}.  The above expression 
(\ref{kkznormint}) for the normalization integral is valid and well-defined, 
only for the case when ${{n+1}\over{n-1}}\neq{4\over\Delta_p}$ and 
${{2n}\over{n-1}}-{4\over\Delta_p}$ is neither zero nor a negative integer.  
These conditions are satisfied by $\Delta_p$ defined in Eq. (\ref{magpbrndef}) 
with $n\geq 2$ and $p\geq 3$.  The normalization integral (\ref{kkznormint}) 
does not have a diverging contribution from $y=0$, where the $p$-brane is 
located, if ${{n+1}\over{n-1}}>{4\over{\Delta_p}}$.  This condition is also 
satisfied by $\Delta_p$ defined in Eq. (\ref{magpbrndef}) with $n\geq 2$ 
and $p\geq 3$.  Note, this condition is essential in normalization of 
the KK zero mode, since we are not allowed to truncate the transverse space 
to exclude the $y=0$ region, where the $p$-brane is located.  However, the 
normalization integral (\ref{kkznormint}) always has a diverging contribution 
from $y=\infty$ for any values of $n$ and $\Delta_p$.  So, one has to truncate 
the transverse space (namely, take the integration interval in the 
normalization integral to be $0\leq y\leq y_0$ with $y_0<\infty$), if one 
wishes to normalize the KK zero mode.  A possible scenario with such truncated 
integration interval is the one proposed in Ref. \cite{groj}, where the 
{\it jump brane} (identifiable as the {\it Planck brane}) at a finite distance 
from the $p$-brane is introduced through the T-dualization of the transverse 
space.  

Since the dilatonic $p$-brane solution (\ref{magpbrnsol}) is asymptotically 
flat, it may be possible to reproduce the $(p+1)$-dimensional gravity in the 
intermediate distance region on the worldvolume of the $p$-brane through the 
massive KK modes by applying the mechanism proposed in Ref. \cite{quas2}.  
However, we shall not pursue this direction, since its validity for describing 
our world is rather controversial \cite{quas3,quas4,witt,quas5,quas6,quas7}.   
Also, study of massive graviton KK modes in the bulk of a $p$-brane is a lot 
more complicated than the domain wall case.  The coordinate transformation 
that brings the $(p+2)$-dimensional part of the $p$-brane metric (with the 
coordinates $(x^{\mu},|{\bf y}|)$) into the conformally flat form, in which 
the equation satisfied by the metric fluctuation takes the Schr\"odinger 
equation form, involves a special function of the coordinate $|{\bf y}|$, 
which cannot be inverted to re-express the metric in new coordinates.

\section{Delocalized $p$-Branes}

In the previous section, we saw that the KK zero mode of graviton in 
the bulk of a fully localized dilatonic $p$-brane (with asymptotically 
flat spacetime) is not normalizable, if the transverse space of the 
$p$-brane is of infinite size.  In this section, we attempt to normalize 
the graviton KK zero mode by delocalizing the $p$-brane along its 
transverse directions except one.  Delocalization is achieved by 
constructing dense periodically arrayed multi-center $p$-branes along 
the transverse directions to be delocalized.  The delocalization process 
can also be regarded as first placing $p$-branes at equivalent points 
of the compactification lattices and then taking the limit of very 
small compactification manifold.  Therefore, the spacetime of such 
delocalized $p$-brane can also be thought of as the product of a 
$(p+1)$-dimensional domain wall spacetime and an $n$-dimensional compact 
space.  

The delocalized $p$-brane solution has the following form:
\begin{equation}
G_{MN}dx^Mdx^N=H^{-{{4(n-1)}\over{(p+n)\Delta_p}}}_p\left[-dt^2+d{\bf x}
\cdot d{\bf x}\right]+H^{{4(p+1)}\over{(p+n)\Delta_p}}_p\left[d\tilde{y}^2
+d\tilde{s}^2_n\right],
\label{delocsol}
\end{equation}
where $\tilde{y}$ is one of the transverse coordinates ${\bf y}$ and 
$d\tilde{s}^2_n$ is the metric on an $n$-dimensional compact manifold 
${\cal K}_n$ upon which the $p$-brane is delocalized and the harmonic 
function is given by $H_p=1+\tilde{Q}_p|\tilde{y}|$.  If $\tilde{Q}_P<0$, 
the location $|\tilde{y}|=-\tilde{Q}^{-1}_p$, where the harmonic function 
$H_p$ vanishes, corresponds to the curvature singularity.  When $\tilde{Q}_p
>0$, there is no singularity except at $y=0$, where the delocalized $p$-brane 
is located.  

For the purpose of studying the delocalized $p$-brane solution in relation to 
the RS type scenario, it is convenient to transform the transverse coordinate 
$\tilde{y}$ so that the $(p+2)$-dimensional part of the metric (with the 
coordinates $(x^{\mu},\tilde{y})$) takes the standard form of the RS model 
with the warp factor ${\cal W}$ or the conformally flat form with the 
conformal factor ${\cal C}$.  The solutions in the new coordinates are given by
\begin{equation}
G_{MN}dx^Mdx^N={\cal W}\left[-dt^2+d{\bf x}\cdot d{\bf x}\right]+
dy^2+{\cal W}^{-{{p+1}\over{n-1}}}d\tilde{s}^2_n,
\label{warpsol}
\end{equation}
where the warp factor is given by
\begin{eqnarray}
{\cal W}(y)&=&\left[1\pm{{2(p+1)+(p+n)\Delta_p}\over{(p+n)\Delta_p}}
\tilde{Q}_p|y|\right]^{-{{4(n-1)}\over{2(p+1)+(p+n)\Delta_p}}}
\cr
&=&\left[1\pm{{(p+n)^2a^2_p+4(p+1)n}\over{(p+n)^2a^2_p+4(p+1)(n-1)}}
\tilde{Q}_p|y|\right]^{-{{8(n-1)}\over{(p+n)^2a^2_p+4(p+1)n}}},
\label{warpfac}
\end{eqnarray}
and 
\begin{equation}
G_{MN}dx^Mdx^N={\cal C}\left[-dt^2+d{\bf x}\cdot d{\bf x}+dz^2\right]
+{\cal C}^{-{{p+1}\over{n-1}}}d\tilde{s}^2_n,
\label{confsol}
\end{equation}
where the conformal factor is given by
\begin{equation}
{\cal C}(z)=\left[1\pm{{\Delta_p+2}\over{\Delta_p}}\tilde{Q}_p|z|
\right]^{-{{4(n-1)}\over{(p+n)(\Delta_p+2)}}}.
\label{confact}
\end{equation}
The expressions for dilaton $\phi$ and the $n$-form field strength $F_n$ in 
terms of the warp factor or the conformal factor can be obtained by using the 
relation ${\cal W}(y)=H^{-{{4(n-1)}\over{(p+n)\Delta_p}}}_p(\tilde{y})=
{\cal C}(z)$.  

We notice the following important difference of the warp factor 
(\ref{warpfac}) for the delocalized $p$-brane from that (\ref{warpdil}) 
of the dilatonic domain wall (of codimension one).  
In the case of the dilatonic domain walls, the exponent in the warp factor 
(\ref{warpdil}) is always positive.  So, one has to choose the sign 
$\pm$ in Eq. (\ref{defk}) such that the warp factor has zeros at finite 
nonzero $y$ on both sides of the wall, if one wants to have the decreasing 
warp factor.  The side-effect of such choice of sign is the naked 
singularities at the positions where the warp factor vanishes.  On the other 
hand, in the case of the delocalized $p$-branes, the exponent in the warp 
factor (\ref{warpfac}) is always negative.  So, in order to have a decreasing 
warp factor on both sides $y>0$ and $y<0$, one has to choose the sign $\pm$ 
in Eq. (\ref{warpfac}) such that the term in the square bracket does not have 
a root at finite nonzero $y$.  With such choice of the sign, the delocalized 
$p$-brane solution does not have problematic naked singularities.  With a 
choice of the sign $\pm$ such that the term in the square bracket has a root 
at a finite nonzero $y$ on both sides of the brane, the warp factor ${\cal W}$ 
increases, asymptotically approaching infinity as the roots are reached.   
Such roots correspond to the curvature singularities.  However, as we will 
see in the following, unlike the case of dilatonic domain walls of 
codimension one, the graviton KK zero mode in the bulk of delocalized 
$p$-brane is normalizable when the warp factor ${\cal W}$ {\it increases} on 
both sides of the $p$-brane.  

To study the KK modes of the bulk graviton, we consider the following 
small fluctuation around the delocalized $p$-brane metric:
\begin{equation}
G_{MN}dx^Mdx^N={\cal C}\left[\left(\eta_{\mu\nu}+h_{\mu\nu}\right)
dx^{\mu}dx^{\nu}+dz^2\right]+{\cal C}^{-{{p+1}\over{n-1}}}d\tilde{s}^2_n,
\label{metfluc}
\end{equation}
where the metric perturbation $h_{\mu\nu}(x^{\rho},z)$, which is 
taken to be independent of the coordinates of the $n$-dimensional compact 
space ${\cal K}_n$ with the metric $d\tilde{s}^2_n$, is assumed to satisfy 
the transverse traceless gauge condition $h^{\mu}_{\ \mu}=0=\partial^{\mu}
h_{\mu\nu}$.   Then, the $(\mu,\nu)$-component of the Einstein's equations 
is approximated, to the first order in $h_{\mu\nu}$, to
\begin{equation}
\left[\Box_x+\partial^2_z-{{p+n}\over{2(n-1)}}{{\partial_z{\cal C}}\over
{\cal C}}\partial_z\right]h_{\mu\nu}=0,
\label{lineineq}
\end{equation}
where $\Box_x\equiv\eta^{\mu\nu}\partial_{\mu}\partial_{\nu}$.  To 
consider the KK mode with mass $m_{\alpha}$, only, we decompose $h_{\mu\nu}$ 
as $h_{\mu\nu}(x^{\rho},z)=\hat{h}^{(\alpha)}_{\mu\nu}(x^{\rho})f_{\alpha}(z)$ 
and require $\hat{h}^{(\alpha)}_{\mu\nu}$ to satisfy $\Box_x
\hat{h}^{(\alpha)}_{\mu\nu}=m^2_{\alpha}\hat{h}^{(\alpha)}_{\mu\nu}$.  Then, 
the linearized Einstein's equations (\ref{lineineq}) reduce to the following 
form:
\begin{equation}
\left[\partial^2_z-{{p+n}\over{2(n-1)}}{{\partial_z{\cal C}}\over
{\cal C}}\partial_z+m^2_{\alpha}\right]f_{\alpha}(z)=0.
\label{lineineq2}
\end{equation}
This equation can be brought to the following form of the Sturm-Liouville
equation:
\begin{equation}
\partial_z\left[{\cal C}^{-{{p+n}\over{2(n-1)}}}\partial_zf_{\alpha}\right]+
m^2_{\alpha}{\cal C}^{-{{p+n}\over{2(n-1)}}}f_{\alpha}=0,
\label{gravkksleq}
\end{equation}
from which we see that the KK modes $f_{\alpha}(z)$ with different masses 
$m_{\alpha}$ are orthogonalized with respect to the weighting function 
$w(z)={\cal C}^{-{{p+n}\over{2(n-1)}}}$.  Had we used the transverse 
coordinate $y$, instead of $z$, the equation satisfied by the KK mode 
$f_{\alpha}$ would have taken the following form:
\begin{equation}
\partial_y\left[{\cal W}^{-{{p+1}\over{2(n-1)}}}\partial_yf_{\alpha}\right]+
m^2_{\alpha}{\cal W}^{-{{p+2n-1}\over{2(n-1)}}}f_{\alpha}=0,
\label{gravkksleq2}
\end{equation}
from which we know that the KK modes $f_{\alpha}(y)$ are orthogonalized with 
respect to the weighting function $w(y)={\cal W}^{-{{p+2n-1}\over{2(n-1)}}}$.  

In terms of a new $z$-dependent function defined as $\tilde{f}_{\alpha}\equiv 
{\cal C}^{-{{p+n}\over{4(n-1)}}}f_{\alpha}$, Eq. (\ref{gravkksleq}) takes 
the following form of the Schr\"odinger equation:
\begin{equation}
-{{d^2\tilde{f}_{\alpha}}\over{dz^2}}+V(z)\tilde{f}_{\alpha}=
m^2_{\alpha}\tilde{f}_{\alpha},
\label{schrodeq}
\end{equation}
with the potential
\begin{eqnarray}
V(z)&=&{{p+n}\over{16(n-1)^2}}\left[(p+5n-4)\left({{{\cal C}^{\prime}}\over
{\cal C}}\right)^2-4(n-1){{{\cal C}^{\prime\prime}}\over{\cal C}}\right]
\cr
&=&-{{(\Delta_p+1)\tilde{Q}^2_p}\over{\Delta^2_p}}{1\over{(1\pm{{\Delta_p+2}
\over{\Delta_p}}\tilde{Q}_p|z|)^2}}\pm{{2\tilde{Q}_p}\over{\Delta_p}}\delta(z),
\label{schrodpot}
\end{eqnarray}
where the order in the sign $\pm$ is the same as that in Eq. (\ref{confact}).  
The zero mode solution $\tilde{f}_0$ (corresponding to the zero KK mass 
$m_0=0$) to the Schr\"odinger equation satisfying the boundary condition 
$\tilde{f}^{\prime}_0(0^+)-\tilde{f}^{\prime}_0(0^-)=\pm{{2\tilde{Q}_p}\over
{\Delta_p}}\tilde{f}_0(0)$ is $\tilde{f}_0\sim (1\pm{{\Delta_p+2}\over
{\Delta_p}}\tilde{Q}_p|z|)^{1\over{\Delta_p+2}}$.  So, the KK zero mode is 
independent of $z$: $f_0(z)={\cal C}^{{p+n}\over{4(n-1)}}\tilde{f}_0=
{\rm constant}$.  By calculating the normalization integration $\int dy\,f^2_0
w(y)=\int dy\,f^2_0{\cal W}^{-{{p+2n-1}\over{2(n-1)}}}$, one can see that 
the KK zero mode $f_0$ is normalizable when the sign $\pm$ in the warp 
factor (\ref{warpfac}) is chosen so that the warp factor {\it increases} on 
both sides of the $p$-brane, in which case there are curvature singularities 
at finite non-zero $y$.  With another choice of the sign, even if the warp 
factor decreases, the graviton KK zero mode is not normalizable.

Recently, it is observed \cite{mvv,grs} that free massive particles in the 
bulk of (non-dilatonic) domain wall with exponentially decreasing warp 
factor of RS are repelled by the domain wall into the extra spatial 
direction, whereas those in the domain wall with exponentially increasing 
warp factor are attracted towards the domain wall.  This undesirable feature 
of the RS domain wall calls for need to find mechanism of trapping massive 
matter within the domain wall so that matter in our world is not lost into 
the extra dimension.  

Generally, one can show that free massive particles in the bulk of domain 
wall with decreasing [increasing] warp factor are repelled from [attracted 
towards] the domain wall, as follows.  We consider the following general 
form of metric with the warp factor ${\cal W}(y)$:
\begin{equation}
g_{\mu\nu}dx^{\mu}dx^{\nu}={\cal W}(y)\left[-dt^2+d{\bf x}\cdot d{\bf x}
\right]+dy^2.
\label{genwapmet}
\end{equation}
By contracting the Killing vectors $\partial/\partial t$ and 
$\partial/\partial x^i$ of the metric (\ref{genwapmet}) with the 
velocity $U^{\mu}=dx^{\mu}/d\lambda$ of a free test particle along 
its geodesic path $x^{\mu}(\lambda)$ parameterized by an affine 
parameter $\lambda$, one obtains the following constants of motion for 
the test particle:
\begin{eqnarray}
E&=&-g_{\mu\nu}\left({{\partial}\over{\partial t}}\right)^{\mu}U^{\nu}=
-g_{tt}{{dt}\over{d\lambda}}={\cal W}(y){{dt}\over{d\lambda}},
\cr
p^i&=&g_{\mu\nu}\left({{\partial}\over{\partial x^i}}\right)^{\mu}
U^{\nu}=g_{x^ix^i}{{dx^i}\over{d\lambda}}={\cal W}(y){{dx^i}\over
{d\lambda}},
\label{cnstmtn}
\end{eqnarray}
which can be thought of as the energy and linear momentum for massless 
particles and the energy and linear momentum per unit mass for massive 
particles.  In addition, metric compatibility for the geodesic motion 
implies 
\begin{equation}
\epsilon=-g_{\mu\nu}{{dx^{\mu}}\over{d\lambda}}{{dx^{\nu}}\over{d\lambda}}=
{\cal W}\left[\left({{dt}\over{d\lambda}}\right)^2-{{d{\bf x}}\over
{d\lambda}}\cdot{{d{\bf x}}\over{d\lambda}}\right]-\left({{dy}\over
{d\lambda}}\right)^2,
\label{geomtn}
\end{equation}
where $\epsilon=1,0$ respectively for a massive and a massless test 
particle.   By making use of Eq. (\ref{cnstmtn}), one can bring Eq. 
(\ref{geomtn}) into the following form:
\begin{equation}
\epsilon=(E^2-{\bf p}\cdot{\bf p}){\cal W}^{-1}-\left({{dy}\over{d\lambda}}
\right)^2,
\label{geomtn2}
\end{equation}
where ${\bf p}=(p^i)$.  Note, the test particles do not feel any force 
along the ${\bf x}$-direction, because the metric (\ref{genwapmet}) does not 
depend on ${\bf x}$.  So, it is possible to consider the geodesic motion with 
$d{\bf x}/d\lambda=0$, or one can just move to a frame in which a massive 
test particle moves along the $y$-direction by using the boost 
invariance of the metric along the ${\bf x}$-direction.  In this case, the 
velocity of a massive test particle along the $y$-direction is given by
\begin{equation}
{{dy}\over{d\lambda}}=\pm\sqrt{{{E^2}\over{\cal W}}-1}.
\label{velydirec}
\end{equation}
Note, this equation is valid also for the motion of a massive test particle 
along the $y$-direction in the spacetime with the metric (\ref{warpsol}), 
since the massive test particle does not feel any force along the isometry 
directions of the $n$-dimensional compact manifold ${\cal K}_n$ if we, for 
example, take ${\cal K}_n=T^n$.  For a decreasing warp factor ${\cal W}$, 
the velocity $dy/d\lambda$ of a free massive particle ($\epsilon=1$) along 
the $y$-direction increases as it moves away from the domain wall, 
implying that the domain wall repels a massive particle.  In the case 
of a massless test particle ($\epsilon=0$), as can be seen from Eq. 
(\ref{geomtn2}), its motion can be confined within the worldvolume directions 
(since $E^2-{\bf p}\cdot{\bf p}=0$ for such motion), but it will also be 
repelled by the domain wall once it has non-zero velocity along the 
$y$-direction.  For a increasing warp factor, the velocity of a free massive 
particle ($\epsilon=1$) along the $y$-direction decreases as it moves away 
from the domain wall and then the particle reflects back to the domain wall 
at the turning point $y=y_0$, given by ${\cal W}(y_0)=E^2$, implying that the 
domain wall attracts the massive particle.  However, the massless particle 
($\epsilon=0$) can move along the worldvolume directions, but it will 
continue to move alway from the domain wall with its velocity asymptotically 
approaching zero, if ${\cal W}(y)\to\infty$ as one moves away from the domain 
wall, once its velocity has nonzero component along the $y$-direction.  Note, 
for massless particles it is $dy/d\lambda$ that is changing, whereas the 
speed of light remains constant.  

It is therefore quite problematic for the dilatonic domain wall 
(\ref{warpdwmet}) with decreasing warp factor, because massive test particles 
will be repelled away from the domain wall and hit the singularity at 
$|y|=-K^{-1}$.  On the other hand, in the bulk background of the delocalized 
$p$-brane (\ref{warpsol}) with increasing warp factor (\ref{warpfac}), a 
massive test particle with a finite energy $E$ will always be reflected back 
to the domain wall before it reaches the singularities, since the warp factor 
${\cal W}$ is monotonically increasing, approaching infinity as the 
singularities are reached.  So, even if the delocalized $p$-brane with 
increasing warp factor suffers from singularities at finite non-zero $y$, 
massive matter is trapped within the $p$-brane by gravitational force and 
can never reach the singularities.  One can think of a dilatonic domain wall 
as being compactified from a (intersecting) $p$-brane in higher dimensions.  
Namely, starting from a dilatonic $p$-brane solution (\ref{magpbrnsol}), 
constructing a dense periodic array of parallel $p$-branes along $n$ 
transverse directions to obtain the delocalized $p$-brane (\ref{delocsol}), 
and then dimensionally reducing along the $n$ delocalized transverse 
directions, one obtains the dilatonic domain wall (\ref{dwsol}) in $D=p+2$ 
dimensions with the following dilaton coupling parameter:
\begin{equation}
a^2={{(p+n)a^2_p}\over{p}}+{{4(p+1)^2n}\over{p^2(p+n)}}.
\label{dilpara}
\end{equation}
So, if we take the spacetime to be the product of the domain wall spacetime 
and a compact manifold ${\cal K}_n$, instead of taking it as just the 
spacetime of the domain wall, we do not face the problem of test particles 
repelled to the extra spatial direction and hitting naked singularities of 
the dilatonic domain wall with decreasing warp factor.   

From this picture of dilatonic domain walls, we see that the naked 
singularities of dilatonic domain walls with the decreasing warp factor 
can be regarded as a consequence of dimensionally reducing a non-BPS 
$p$-brane.  Namely, for the BPS case, the parameter $Q_p$ of the $p$-brane 
solution (\ref{magpbrnsol}) is positive and its corresponding delocalized 
$p$-brane solution (\ref{delocsol}), therefore, has positive $\tilde{Q}_p$.  
So, the corresponding dimensionally reduced dilatonic domain wall solution 
(\ref{dwharmf}) will have positive $Q$, thereby having no naked 
singularities at finite nonzero $x_{D-1}$.  And the same has to be true 
for the domain wall solution (\ref{warpdwmet}) in different coordinates    
\footnote{This requirement of the similar singularity structure in 
different coordinates can be used to fix the sign ambiguity in the 
coordinate transformation from $x_{D-1}$ to $y$, which brings the 
metric (\ref{dwsol}) to the form (\ref{warpdwmet}).  This sign 
ambiguity in the coordinate transformation (resulting from $H^{{2(D-1)}
\over{(D-2)\Delta}}dx_{D-1}=\pm dy$) appears as the ambiguity of the 
choice of sign $\pm$ in Eq. (\ref{defk}).}.  In the case of a non-BPS 
$p$-brane, $Q_p<0$ and therefore $Q<0$ in the corresponding dilatonic 
domain wall solution, so there are naked singularities away from the wall.


\begin{thebibliography} {99}
\small
\parskip=0pt plus 2pt


\bibitem{add} N. Arkani-Hamed, S. Dimopoulos and G. Dvali, 
``The hierarchy problem and new dimensions at a millimeter,'' 
Phys. Lett.  {\bf B429} (1998) 263, hep-ph/9803315.

\bibitem{aadd} I. Antoniadis, N. Arkani-Hamed, S. Dimopoulos and G. Dvali, 
``New dimensions at a millimeter to a Fermi and superstrings at a TeV,'' 
Phys. Lett. {\bf B436} (1998) 257, hep-ph/9804398.

\bibitem{ran1} L. Randall and R. Sundrum, ``A large mass hierarchy from a 
small extra dimension,'' Phys. Rev. Lett. {\bf 83} (1999) 3370, 
hep-ph/9905221.

\bibitem{ran2} L. Randall and R. Sundrum, ``An alternative to 
compactification,'' Phys. Rev. Lett. {\bf 83} (1999) 4690, hep-th/9906064.

\bibitem{ran3} J. Lykken and L. Randall, ``The shape of gravity,'' 
hep-th/9908076.

\bibitem{bc} K. Behrndt and M. Cveti\v c, ``Supersymmetric domain wall world 
from D = 5 simple gauged supergravity,'' hep-th/9909058.

\bibitem{st} K. Skenderis and P.K. Townsend, ``Gravitational stability and 
renormalization-group flow,'' Phys. Lett. {\bf B468} (1999) 46, 
hep-th/9909070.

\bibitem{cg} A. Chamblin and G.W. Gibbons, ``Supergravity on the brane,'' 
hep-th/9909130.

\bibitem{kls} R. Kallosh, A. Linde and M. Shmakova, ``Supersymmetric multiple 
basin attractors,'' JHEP {\bf 9911} (1999) 010, hep-th/9910021.

\bibitem{bbs} I. Bakas, A. Brandhuber and K. Sfetsos, ``Domain walls of 
gauged supergravity, M-branes, and algebraic curves,'' hep-th/9912132.

\bibitem{clp} M. Cveti\v c, H. Lu and C.N. Pope, ``Domain walls and massive 
gauged supergravity potentials,'' hep-th/0001002.

\bibitem{kl} R. Kallosh and A. Linde, ``Supersymmetry and the brane world,'' 
JHEP {\bf 0002} (2000) 005, hep-th/0001071.

\bibitem{bc2} K. Behrndt and M. Cveti\v c, ``Anti-deSitter vacua of gauged 
supergravities with 8 supercharges,'' hep-th/0001159.

\bibitem{bbs2} I. Bakas, A. Brandhuber and K. Sfetsos, ``Riemann surfaces and 
Schroedinger potentials of gauged supergravity,'' hep-th/0002092.

\bibitem{abn} R. Altendorfer, J. Bagger and D. Nemeschansky, ``Supersymmetric 
Randall-Sundrum scenario,'' hep-th/0003117.

\bibitem{bs} A. Brandhuber and K. Sfetsos, ``Non-standard compactifications 
with mass gaps and Newton's law,'' JHEP {\bf 9910} (1999) 013, hep-th/9908116.

\bibitem{youm1} D. Youm, ``Solitons in brane worlds,'' hep-th/9911218.

\bibitem{mvv} W. M\"uck, K.S. Viswanathan and I.V. Volovich, 
``Geodesics and Newton's law in brane backgrounds,'' hep-th/0002132.

\bibitem{grs} R. Gregory, V.A. Rubakov and S.M. Sibiryakov, 
``Brane worlds: The gravity of escaping matter,'' hep-th/0003109.

\bibitem{stfdw} S. Kachru, M. Schulz and E. Silverstein, 
``Self-tuning flat domain walls in 5d gravity and string theory,'' 
hep-th/0001206.

\bibitem{gjs} C. Gomez, B. Janssen and P. Silva, ``Dilatonic Randall-Sundrum 
theory and renormalization group,'' hep-th/0002042.

\bibitem{youm3} D. Youm, ``Bulk fields in dilatonic and self-tuning flat 
domain walls,'' hep-th/0002147.

\bibitem{cceh} C. Csaki, J. Erlich, T.J. Hollowood and Y. Shirman, 
``Universal aspects of gravity localized on thick branes,'' hep-th/0001033.

\bibitem{groj} C. Grojean, ``$T$ self-dual transverse space and gravity 
trapping,'' hep-th/0002130.

\bibitem{quas2} R. Gregory, V.A. Rubakov and S.M. Sibiryakov, ``Opening up 
extra dimensions at ultra-large scales,'' hep-th/0002072.

\bibitem{quas3} C. Cs\' aki, J. Erlich and T.J. Hollowood, 
``Quasi-localization of gravity by resonant modes,'' hep-th/0002161.

\bibitem{quas4} G. Dvali, G. Gabadadze and M. Porrati, ``Metastable gravitons 
and infinite volume extra dimensions,'' hep-th/0002190.

\bibitem{witt}  E. Witten, ``The cosmological constant from the viewpoint 
of string theory,'' hep-ph/0002297.

\bibitem{quas5} C. Cs\' aki, J. Erlich and T.J. Hollowood, ``Graviton 
propagators, brane bending and bending of light in theories with 
quasi-localized gravity,'' hep-th/0003020.

\bibitem{quas6} R. Gregory, V.A. Rubakov and S.M. Sibiryakov, ``Gravity and 
antigravity in a brane world with metastable gravitons,'' hep-th/0003045.

\bibitem{quas7} G. Dvali, G. Gabadadze and M. Porrati, ``A comment on brane 
bending and ghosts in theories with infinite extra  dimensions,'' 
hep-th/0003054.

\end{thebibliography}
\end{document}